# Imprint of transition metal d-orbitals on graphene Dirac cone

Qin Zhou, Sinisa Coh, Marvin L. Cohen, Steven G. Louie, and A. Zettl

Department of Physics, University of California at Berkeley, and Materials Sciences Division, Lawrence Berkeley National Laboratory, Berkeley, California 94720, USA

## Abstract:

We investigate the influence of $SiO_2$, Au, Ag, Cu, and Pt substrates on the Raman spectrum of graphene. Experiments reveal particularly strong modifications to the intensity, position, width, and shape of the Raman signal of graphene on platinum, compared to that of suspended graphene. The modifications strongly depend on the relative orientation of the graphene and platinum lattices. These observations are theoretically investigated and shown to originate from hybridization of electronic states in graphene and d-orbitals in platinum. It is expected that, quite generally, hybridization between graphene and any material with d-orbitals near the Fermi level will result in an imprint on the graphene Dirac cone which depends sensitively on the relative orientation of the respective lattices.

Raman spectroscopy has proven to be particularly useful in elucidating the vibrational phonon structure of low dimensional solids such as fullerenes, nanotubes, and graphene [1-7]. In graphene, Raman also serves as a convenient, relatively local (~1μm size scale) probe of sample layer number and defect concentration for exfoliated samples or those grown directly on metal substrates such as Cu. Recently there has been success in growing relatively flat and large-grain-size, high quality graphene on platinum[8-10]. Interestingly, the Raman signal from graphene on Pt can be orders of magnitude smaller in intensity than that from graphene on Cu or $SiO_2$. This result, surprising in light of the assumed weak van der Waals interaction between graphene and Pt [8, 10-13], has been vaguely attributed to an unspecified "strong platinum-graphene interaction" [9, 14]. Proper identification and understanding of the suppression mechanism is lacking.

We here contrast the Raman signature of suspended graphene, graphene on $SiO_2$, Au, Ag, Cu, and Pt, including single crystal Pt, and floated graphene brought close to a Pt surface. The results for Pt cannot be accounted for by simple substrate screening. Instead, our experiments and theoretical investigation reveal rich physics underlying the Raman spectrum modification. For graphene on Pt, the Raman spectrum reflects the hybridization between graphene Dirac cone states and Pt d-orbitals, where the hybridization is strongly dependent on the in-plane position of the d-orbital relative to the graphene lattice, and on the orbital character. The s- and d-orbitals interact very differently with the Dirac cone because of the specific nature of the graphene band structure. Thus, and rather remarkably, Raman spectroscopy reveals a detailed imprint of the transition metal d-orbitals on the Dirac cone.

Graphene samples used in this work are prepared by chemical vapor deposition (CVD) [8, 15, 16], either grown directly on the substrate of interest (Cu or Pt polycrystalline foils, or Pt (111) single crystal), or transferred post growth from Cu or Pt onto $SiO_2$ (1 mm thick fused silica), Au, Ag, and selected Pt substrates. The suspended graphene sample is prepared following our previous work[17]. Graphene is removed from Cu via conventional etching [15] and from Pt via bubble release [8]. Raman measurements are performed primarily using a laser wavelength of 488nm or 514nm and power 0.9mW, with a nominal integration time of 10s (or extended for weak signals). All graphene samples are verified to have vanishing D peak, indicating high quality [1-3]. Details of sample preparation and laser measurement can be found in the Supplemental Material (SM) [18].

Fig. 1 shows details of the Raman spectrum near 2600 $cm^{-1}$ (2D peak) and 1600 $cm^{-1}$ (G peak) of suspended graphene and graphene on various substrates. The spectra for graphene transferred onto $SiO_2$, Au, or Ag substrates have reduced intensity, but the 2D and G peak positions, width, and 2D/G intensity ratios are similar to those for suspended graphene. The spectrum of graphene grown directly on Cu shows blue shifted G and 2D peaks, indicating compressive strain [19], as previously reported. We find that the spectrum of graphene grown on Pt has dramatically reduced and sometimes substantially shifted 2D and G peaks. Fig. 1 shows one example, where the 2D peak for graphene on Pt is nearly four orders of magnitude smaller than that for suspended graphene, and blue shifted by over 50 $cm^{-1}$. The G peak for the same sample region is reduced by nearly two orders of magnitude. We have also prepared a sample of graphene floated on water, where the graphene is brought close to a bare Pt surface by evaporating the water. The graphene Raman signal is high (and reminiscent of suspended graphene) with a water spacer present but is largely quenched when the graphene touches the Pt (see SM [18] for details).

We first explore electromagnetic screening as a possible cause for the quenching of the Raman signal of graphene on Pt. Electromagnetic screening, also at the heart of Surface-Enhanced Raman Scattering (SERS), arises from reduced optical fields from electromagnetic anti-resonances in the substrates[20]. To a first approximation, the screening factor $S$ can be expressed by the fourth power of the ratio of the total electric field $\mathbf{E}_s$ at the graphene location to the incident excitation field $\mathbf{E}_{in}$ [21]. Employing the Fresnel equations yields

$$S = \left|\frac{\mathbf{E}_s}{\mathbf{E}_{in}}\right|^4 = \left|\frac{\mathbf{E}_{in}+\mathbf{E}_r}{\mathbf{E}_{in}}\right|^4 = \left|1+\frac{1-n}{1+n}\right|^4 = \left|\frac{2}{1+n}\right|^4 \qquad (1)$$

where $n$ is the complex refractive index of the substrate [22]. Eq. (1) predicts S values of 0.43, 0.32, 0.14, 0.13, and 0.03 for $SiO_2$, Au, Ag, Cu, and Pt substrates, respectively. These values are shown in Fig. 1 in black below the respective substrate along with the experimentally derived peak intensities (all normalized to suspended graphene). The observed peak intensities for $SiO_2$, Au, Ag, and Cu substrates can be reasonably well accounted for by screening, but the 2D peak for graphene on Pt is approximately 50 times smaller than expected. Indeed, as we show below, under certain circumstances the 2D peak for graphene on Pt can be virtually undetectable. _ Hence screening is eliminated as the sole cause of Raman signal quenching for graphene on Pt.

We next consider Pauli blocking as a cause of the Raman signal quenching. The G and 2D peaks of graphene result from resonance Raman processes, in which electrons are first excited to the conduction band by the incoming photons and then interact with phonons. Given the large work function difference between graphene and platinum (4.48 and 6.13 eV respectively [23]), graphene could transfer significant charge to Pt, resulting in the electron excitation being Pauli-blocked. However, both theory and experiment indicate a relatively small Fermi level shift of ~0.4 eV relative to the Dirac point[10, 23, 24]. This energy shift is not sufficient to block the optical transition caused by photons with energy up to 2.5 eV (488 nm). We also note that doping or strain could shift the peak position[25, 26], but they have little effect on intensities and peak widths.

Therefore we turn to more in-depth theoretical calculations to clarify the reduction of the Raman 2D signal in graphene on Pt. Fig. 2 shows our band structure calculation[27, 28] of graphene on a Pt slab. The distance between the graphene and the platinum slab is kept at $z$=3.3 A [10, 29], and we consider the three most common relative orientations of the graphene lattice with respect to the Pt (111) surface [30]. The Dirac cone near the Dirac point is strongly hybridized when in contact with the Pt slab. Furthermore, the size and the orbital character (different colors in Fig. 2) of the hybridization gap depend strongly on the relative orientation of the graphene lattice with respect to the Pt lattice (compare orientations A, B, and C in Fig. 2).

From Fig. 2 it is clear that the effect of the Pt hybridization with the graphene Dirac cone is quite complicated. Therefore, for illustration purposes we first work with a simplified model in which graphene states are hybridized with only one metallic (s or d) orbital per graphene unit cell, and vary the metallic orbital position and character. Furthermore, we assume that these metallic orbitals form a flat energy band, so that we can easily tune their energy relative to the Dirac point. We parameterize the hybridization strength between the graphene $p_z$ orbitals and the metallic s and d orbitals using density functional theory. This yields the hybridization matrix element between carbon $p_z$ and Pt orbitals at the same in-plane position (head-to-head) but separated vertically by $z$. It is close to -0.2eV for both s and d orbitals (see Fig. 3d and SM[18] for more detail), but as the in-plane separation between the carbon $p_z$ and metallic orbitals is increased hybridization with strongly anisotropic d-orbitals results in a much faster decay than with the isotropic s-orbitals. In this simplified model an almost negligible hybridization gap is opened by the metallic s-orbitals, while the size of the hybridization gap opened by the metallic d-orbitals is strongly dependent on the orbital d-character and position relative to the graphene lattice. These observations are consistent with those from the calculations shown in Fig. 2, and with previous work [31-34].

Next we compute the graphene Raman G and 2D signals using a simplified model. We neglect the effects of the metallic slab on the graphene phonon frequencies and focus only on the electronic state modifications. Our Raman calculation shows that the hybridization of graphene states with metal orbitals reduces the Raman 2D signal and that this reduction is a direct measure of the hybridization gap size. Furthermore, hybridization also shifts the Raman 2D peak position, increases its width, and introduces new Raman peak substructure (see SM[18] for more detail). Fig. 3 shows the dependence of the Raman 2D signal reduction on the orbital position (panels a, b, c in Fig. 3), orbital character (colors in Fig. 3), and energy alignment relative to the Dirac cone (horizontal axis in Fig. 3). Comparing the effect of s and d orbitals on the 2D signal reduction, we find that s-orbitals have an almost negligible effect, as they open a negligible hybridization gap. Additionally, the effect of d-orbitals is strongly dependent on both d-orbital position and orbital character (for example, the $d_z^2$ orbital has almost no effect if placed on a "top" site as compared to placing it on the "hollow" site.). Therefore, we expect that the Raman 2D signal will be strongly dependent on the relative orientation of graphene lattice with respect to the Pt lattice. Finally, the Raman 2D signal reduction is maximal when the hybridization gap is well matched with the energy of the incoming photons (in the case of Fig. 3 this corresponds to the hybridization gap being near half of the laser photon energy (1.96 eV) below the Dirac point).

Fig. 3 shows that *one* metallic d-orbital per graphene unit cell reduces the Raman 2D intensity at most by a factor of four. Taking into account a more realistic number of d-orbitals per graphene unit cell (~5) and repeating our model calculation for this case, we find that the 2D intensity of graphene on platinum can be reduced up to 20 times.

Unlike the case for the Raman 2D signal, we find almost no influence of hybridization on the Raman G signal intensity. This can be explained by considering the different origin of the Raman G signal compared to the 2D signal [35]. The Raman G signal intensity even in suspended graphene is severely reduced in intensity because of the coherent cancellation between amplitudes of various electron-hole pairs in the Dirac cone. In fact, a perfectly linear Dirac cone dispersion leads to a Raman G signal with vanishing intensity. Therefore, any small imperfections in the band structure (such as trigonal warping), will lead to an incomplete cancellation of the Raman amplitudes, and will thus produce a measurable Raman G signal. This observation also explains why we find a small *increase* in the calculated Raman G signal intensity upon hybridization with metallic orbitals (see SM[18]), as hybridization with metallic orbitals leads to a more incomplete cancellation of the G signal amplitudes.

Our theoretical analysis predicts that the Raman 2D (but not G) peak of graphene on Pt will be highly dependent on the relative orientation between the graphene and Pt lattices. To experimentally obtain a range of different lattice orientations we grow large area graphene on a polycrystalline platinum foil. We also grow single-domain graphene on single crystal Pt (111).

Fig. 4 shows graphene Raman spectra measured at different locations on the polycrystalline Pt substrate and the single crystal Pt substrate (laser wavelength is 488nm and integration time is from 500s to 10,000s; data for other laser wavelengths can be found in SM [18]). For the polycrystalline substrate a small variable shift of the position of the G peak (from 0 to 25 cm$^{-1}$) is found, with nearly constant intensity. The intensity robustness of the G peak is in agreement with the theoretical discussion above. The small shift of the G peak likely originates from an inhomogeneous strain field developed during the cooling process [36, 37] after graphene synthesis. On the other hand, the intensity, width, position, and shape of the Raman 2D peak vary strongly at different sample positions (representing different lattice misorientations), again in agreement with the theoretical calculations . The position of the 2D peak can shift anywhere between -8 and 100 cm$^{-1}$ with respect to the suspended graphene. The width of the 2D peak is between 25 cm$^{-1}$ and 65 cm$^{-1}$, and it likely contains multiple components. The modulation of the 2D peak intensity is even more noticeable, with observed reductions ranging from 75 to over 10,000 [38], depending on position. We note that the reduction factor expected from screening alone is 29.

The lower insets to Fig. 4 display the degree of spatial inhomogeneity for the G and 2D peak for graphene on Pt. The intensity of the G peak is relatively insensitive to position (i.e. lattice misorientation), while the intensity map of the 2D peak reflects directly the regions of different lattice misorientation. Although not shown in Fig. 4, we find, for single-domain graphene on single crystal Pt (111) with the consistent 2 × 2 supercell structure, thatboth the G and 2D peak intensities are homogeneous, as expected (See SM [18] for more experiment details).

**Acknowledgements.** This research was supported in part by the Director, Office of Energy Research, Office of Basic Energy Sciences, Materials Sciences and Engineering Division, of the U.S. Department of Energy under contract DE-AC02-05CH11231 which provided for Raman spectroscopy, theoretical calculations, and postdoctoral assistance; by the National Science Foundation under grant DMR-1206512 which provided for graphene transfer and structural characterization; and by the Office of Naval Research under grant N00014-12-1-1008 which provided for graphene growth. We thank C. Hwang for help with ARPES measurements, A. T. N'Diaye for help with LEED measurements, and H. Rasool for technical assistance.

**FIGURES:**

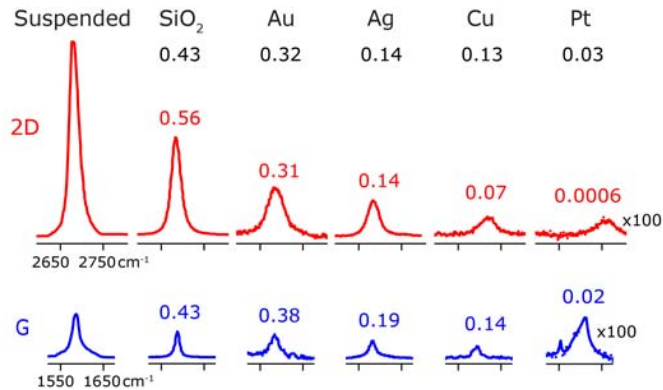

FIG. 1. The Raman signal for graphene on platinum is strongly suppressed compared to suspended graphene, and graphene on $SiO_2$, Au, Ag, and Cu. The x-axes represent the Raman shift and the y-axes represent the amplitude of Raman scattering (linear arb. units). The two distinctive peaks of graphene are the 2D peak at ~ 2700 cm$^{-1}$ (shown in red) and the G peak at ~ 1580 cm$^{-1}$ (shown in blue). The laser wavelength is 514 nm and integration time is 10s for all but Pt. The data for graphene on Pt are acquired by longer integration time (5000s) and the curves are amplified by 100 times for better viewing. The small peak at ~1554 cm$^{-1}$ near the G peak on Pt is from environment oxygen [39]. The numbers (black) below the substrate labels are the predicted peak intensity based only on screening (Eq. 1). The numbers above each curve are the experimentally measured peak intensities. The intensity is defined as the the area covered by the peak and all the intensity values are normalized to the ones from suspended graphene.

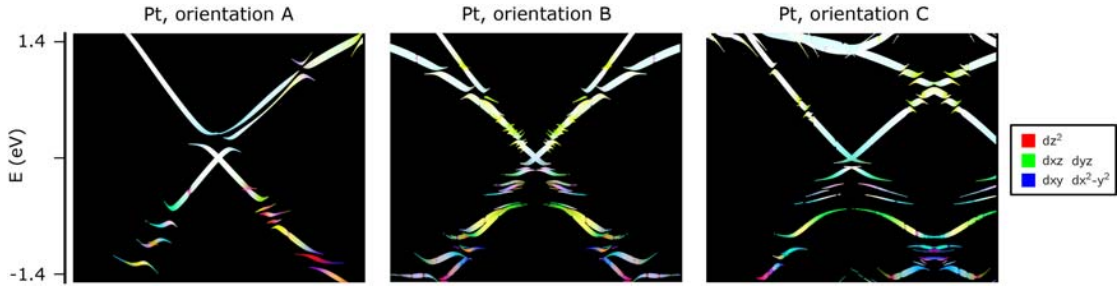

FIG. 2. The graphene Dirac cone is strongly affected by a Ptsubstrate, and varies substantially between the three most common misorientations of the graphene and Pt lattices (orientations A, B, and C correspond to 2x2, 3x3, and 4x4 graphene lattice supercells). The thickness of each line in the plot is proportional to the graphene-like character of the state. Therefore, pure metallic bands are not shown (zero thickness). The color of each line segment is proportional to the mixture of graphene states with different metallic d-orbitals in the top most layer of the Pt slab substrate. Red, green, and blue color component correspond to three different projections of the angular momentum perpendicular to the platinum surface (m=0, +/-1, or +/-2 respectively). Graphene states with no d-character are colored white. The path in reciprocal space for all three misorientations is along the Gamma-K-M line of the primitive graphene Brillouin zone.

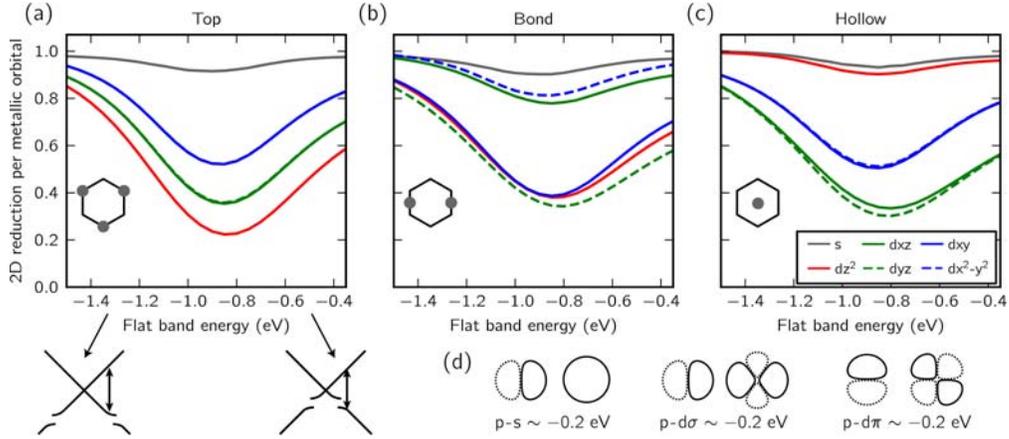

FIG. 3. Reduction of the graphene Raman 2D signal intensity upon hybridization with a metallic flat band (with only one metallic orbital per graphene unit cell). Panels (a), (b), and (c) show reduction of the Raman 2D signal depending on the position of the metallic orbital with respect to the graphene lattice (different panels), orbital character (different line colors and styles), and energy of the metallic band relative to the Dirac cone (horizontal axis). Reduction of the Raman 2D signal is almost negligible for the s orbital, even though its head-to-head matrix element is of the same order of magnitude as for the d orbital (both in sigma and pi orientation, see panel (d)). Furthermore, the effect of d orbital hybridization is strongly dependent on the d-orbital character. The reduction of the Raman 2D signal is largest when the hybridization gap is well matched to the half of the incoming photon energy. The incoming photon energy in this calculation is 1.96 eV, and vertical distance from the metallic d-orbital to the graphene layer is kept constant at 3.3Å. Hybridization parameters are taken from the density functional theory calculation.

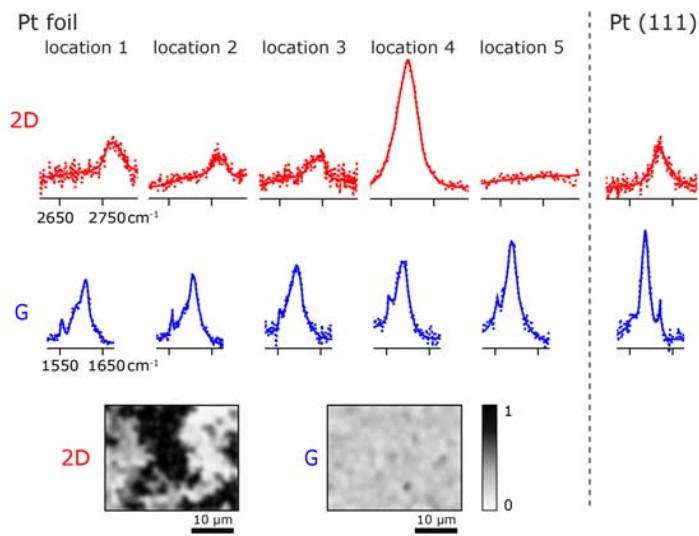

Figure 4. Left:Raman spectra of graphene measured at different locations on a polycrystalline Pt foil.   Laser wavelength: 488 nm.   The intensity maps of a 30 μm × 25 μm region are shown at the bottom.   Right: Raman spectra for graphene on single crystal Pt(111) with 2 × 2 graphene supercell (also shown in Fig. 2).   In this case the intensity for both the 2D and G peaks is found to be uniform across the single domain sample.